\documentclass[
    ,final            
  ]
  {aipproc}

\layoutstyle{8x11double}
\renewcommand\XFMtitleblock{%
 \XFMtitle
 \let\XFMoldpar\par
 \def\par{\XFMoldpar\def\par{\space
            for the VERITAS Collaboration \XFMoldpar}}%
  \XFMauthors
  \let\par\XFMoldpar
  \begin{center}
  \XFMaddresses
\end{center}
  \XFMabstract
  \vspace{5pt}%
  \XFMkeywords
  \XFMclassification
 }

\begin{document}

\title{VERITAS Observations of Extragalactic Non-Blazars}


\author{C. M. Hui}{
  address={Department of Physics, University of Utah, Salt Lake City, UT
    84112, USA},email={cmhui@physics.utah.edu}}

\begin{abstract}
During the 2007/2008 season, VERITAS was used for observations at E>200\,GeV of
several extragalactic non-blazar objects such as galaxy clusters, starburst
and interacting galaxies, dwarf galaxies, and nearby galaxies. In these
proceedings, we present preliminary results from our observations of dwarf
galaxies and M87.   Results from observation of other non-blazar sources are
presented in separate papers in the proceedings. 
\end{abstract}

\maketitle


\section{Introduction}
VERITAS is an array of imaging atmospheric Cherekov telescopes (IACTs) located
at the Fred Lawrence Whipple Observatory on Mount Hopkins in Arizona.  It
consists of four 12m reflectors, each with a camera comprising 499
photomultiplier tubes arranged in a hexagonal lattice covering a field of view
of $3.5^\circ$.  The array is sensitive from 100\,GeV to more than 30\,TeV
and can achieve $1\%$ Crab detection in $\sim$\,50 hours.  For more details see
\citet{Holder08}. 

For the 2007/2008 observing season, $\sim$\,180 hours were allocated for
extragalactic non-blazar sources.  Of these, $\sim$\,20 hours on the Coma
Cluster \cite{coma} and $\sim$\,13 hours on Active Galactic Nuclei (AGNs) near
the Auger hotspots \cite{auger} are described in separate papers in these
proceedings.  In this paper, we will present the preliminary results from our
observation of dwarf galaxies and the non-blazar AGN M87.

\section{Dwarf Galaxies}
A leading candidate for astrophysical dark matter (DM) is a weakly
interacting massive particle (WIMP) with a mass in the range 50 GeV to
more than 10 TeV.  The self-annihilation of WIMPs in astrophysical
regions of high DM density is predicted to generate stable secondary
particles including VHE $\gamma$-rays with energies up to the WIMP mass.
The detection of a $\gamma$-ray source with the unique spectral
characteristics of self-annihilating DM would provide convincing
evidence for the discovery of the particle physics counterpart to
astrophysical DM.

\begin{figure}[t]
  \includegraphics[width=0.48\textwidth]{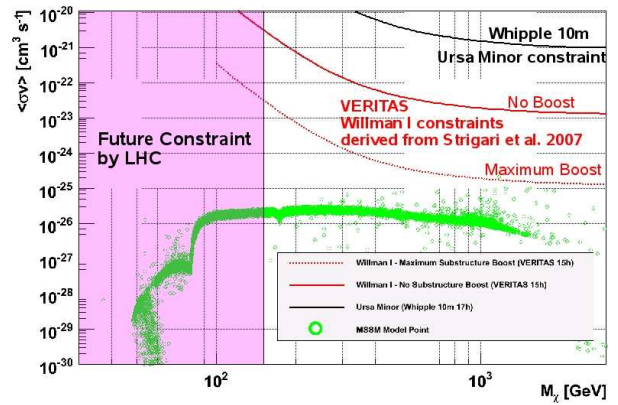}
  \caption{Upper limits on the thermally averaged product of velocity
    and cross section of the WIMP as derived from the 95\% C.L. flux
    upper limits for Willman I.  Upper limits are shown with and
    without a DM substructure boost factor.  Shown as open circles are
    MSSM models that fall within three standard deviations of the
    relic DM density measured in the three-year WMAP data set.}
\label{susy}
\end{figure}

Dwarf spheroidal galaxies of the Local Group are attractive targets to
search for DM annihilations due to their proximity and large DM
content.  Because DM dominates the gravitational potentials of these
systems, their DM distributions can be inferred directly from stellar
kinematics.  Once the DM distribution is known, constraints on the
properties of the WIMP can be derived in the absence of a detection.
VERITAS undertook observations of three dwarf galaxies during the
period 2007-2008: Draco, Ursa Minor, Willman I.  Draco and Ursa Minor
are relatively high surface brightness objects for which large
spectroscopic data sets exist.  Willman I belongs to the group of
faint stellar systems that were recently discovered in the Sloan
Digital Sky Survey \citep{willman05}.  Although Willman I
shares some similarities with tidally disrupted globular clusters, the
extreme mass-to-light ratio inferred from its stellar velocity
dispersion \citep{martin07} suggests that it is a dwarf galaxy.
\citet{strigari08} have projected that this system may have
significantly larger DM annihilation signal than that calculated for
previously known dwarf galaxies.

VERITAS took 20 hours of observation on both Draco and Ursa Minor from
March 2007 to May 2007 and 15 hours of observation on Willman I from
December 2007 to Feburary 2008.  No significant $\gamma$-ray emission were
detected, and the 95\% C.L. upper limits on the integral $\gamma$-ray flux
from each source were set at the level of approximately 1 percent of the Crab
Nebula flux.  Using the DM distribution model presented in \citet{strigari08},
constraints on the thermally averaged product of velocity and cross section of
WIMP ($\left<{\sigma}v\right>$) as a function of its mass were derived from
the $\gamma$-ray flux upper limits for Willman I (see Figure \ref{susy}).  For
comparison with these constraints, models from the framework of the minimal
supersymmetric extension to the standard model (MSSM) were generated with the
DarkSUSY code \citep{gondolo05}. The contribution of DM substructures can
potentially boost the DM annihilation flux with respect to that expected for a
smooth DM halo. The magnitude of this boost depends on the the slope and
cutoff of the subtructure mass function.  \citet{strigari07} estimated the
upper limit for the subtructure enhancement to be $\sim$100.  The upper limits
on $\left<{\sigma}v\right>$ in Figure \ref{susy} are shown for the cases of no
enhancement and maximal enhancement from DM substructure.

\section{M87}
M87 is a nearby giant elliptical galaxy about 16 Mpc away near the center of
the Virgo cluster and has been observed in energy range from radio to TeV
$\gamma$-rays.  Its core is an active galactic nucleus (AGN) powered by a
supermassive black hole and its jet, resolved in radio, optical, and X-ray, is
oriented within $19^\circ$ of the observer's line of sight, derived from
superluminal motion observed in optical \citep{biretta99} and within
$26^\circ$ from radio observation \citep{cheung07}.  Hence M87 is described as
a misaligned BL Lac \citep{reimer04} \citep{tsvetanov97}. 

TeV emission from M87 was first reported by the HEGRA collaboration from their
1998-1999 observations \cite{hegra03}.  This was confirmed by the
H.E.S.S. collaboration \cite{hess06}, which additionally reported year-scale
flux variability and fast (2-day scale) variability during a high state of
$\gamma$-ray activity in 2005.  At the same period, Chandra recorded the
compact knot HST-1, $\sim$\,0.8" from the core, at $\sim$\,50 times its
observed intensity in 2000 \cite{harris07}. While the rapid variability in TeV
emission suggests a compact source region, most likely close to the black
hole, it remains unclear whether the TeV emission comes from the core or the
knot HST-1. 

In 2007 VERITAS confirmed TeV emission above 250\,GeV from M87 but at a lower
flux than what was reported by H.E.S.S. in 2005, and no rapid variability was
seen \cite{veritas08}.  In the same paper, a year-scale correlation was
suggested between $\gamma$-ray flux recorded over the past decade and X-ray
flux in the $2-10$\,keV energy range recorded by ASM/RXTE.  From this
correlation and the non-detection from ASM of the 2005 flare in HST-1 knot
observed by Chandra, the core was suggested to be most likely the TeV emission
site.  However, such $\gamma$/X-ray correlation was not seen with Chandra data
in the $0.2-6$\,keV range.  Since ASM/RXTE and Chandra have an overlap between
$2-6$\,keV, to explain the observed correlation the spectrum would need to
harden significantly within a very limited energy range ($6-10$\,keV) which
seems rather unphysical.  A detailed examination of the TeV and ASM/RXTE X-ray
correlation was presented in \citet{cheung08}, where the possibility of source
contamination from a nearby galaxy and the annual modulation of M87 data from
RXTE were investigated.

The day-scale variability was confirmed by the MAGIC collaboration
\cite{magic08} which reported a 13-day flare during the 2008
H.E.S.S./MAGIC/VERITAS joint monitoring campaign \cite{matthias}.  VERITAS
received a trigger alert from the MAGIC collaboration during that time and
we detected another flare and sent out a trigger alert.  The preliminary
results from VERITAS during this campaign are presented here. 

\begin{figure}[b]
  \includegraphics[width=0.48\textwidth]{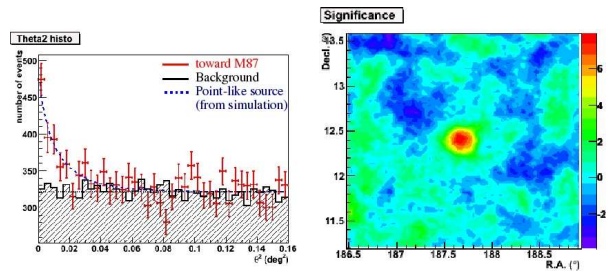}
  \caption{M87 $\theta^2$ distribution and significance map for 2008
    observations.} 
\label{M87signi}
\end{figure}

\begin{figure}[t]
  \includegraphics[width=0.5\textwidth]{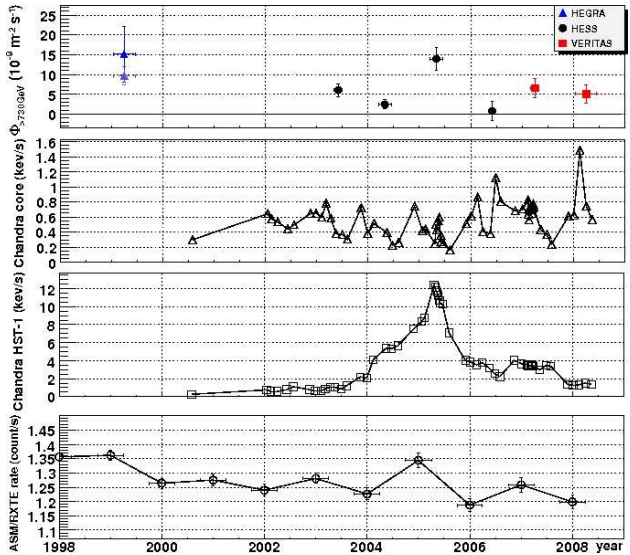}
  \caption{M87 $\gamma$-ray and X-ray lightcurves over the past decade.
    Chandra X-ray data is provided by D. E. Harris via private communication,
    ASM/RXTE quick-look result is provided by the ASM/RXTE team on the web.}
\label{M87year}
\end{figure}

M87 was observed with VERITAS for over 43 hours between December 2007 and May
2008 at a range of zenith angle from $19^\circ$ to $36^\circ$.  All the
observations were performed with 4 telescopes in wobble mode.  After
eliminating bad weather observations and unstable trigger rate data, 41 hours
of quality data ($95\%$ of the whole dataset) yielded a $7.08 \sigma$
detection, with a time-averaged excess rate $0.14 \pm 0.02$
$\gamma\,min^{-1}$, corresponding to $2.1\%$ Crab rate observed at similar
zenith angles. 

The 2008 VERITAS data showed an indication of fast variability in Feburary,
comparable to what was reported by H.E.S.S. in 2005 \cite{hess06} (details in
forthcoming publication).  Preliminary analysis shown that during the 3 nights
with elevated excess rate, the averaged $\gamma$-ray rate corresponded to
$9\%$ Crab rate, while the rest of the dataset averaged to $1\%$ Crab rate.
At the time of the flare, Chandra measured the core activity at a historical
maximum, $\sim\,5$ times its intensity in 2000, while the nearby compact knot
HST-1 was in a quiet state, with a lower flux than the core (see figure
\ref{M87year}).    

\begin{figure}
  \includegraphics[width=0.4\textwidth]{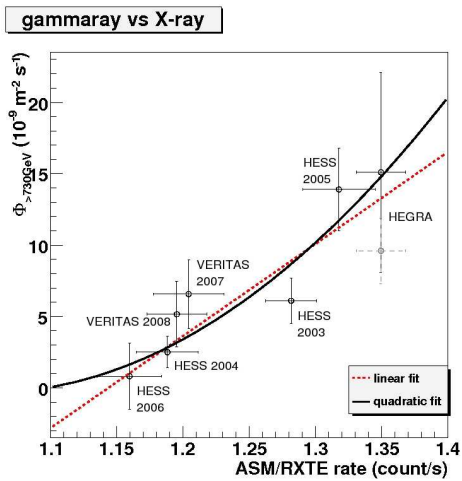}
  \caption{M87 $\gamma$-ray and X-ray correlation over the past decade.  X-ray
    data is the 5-month average (January through May) of quick-look result
    provided by the ASM/RXTE team on the web.}  
\label{M87corr}
\end{figure}

We have also regenerated the $\gamma$/X-ray correlation published in
\citet{veritas08} (figure \ref{M87corr}) with the updated ASM results from the
ASM/RXTE website and this year's data from VERITAS.  The 2007 VERITAS flux
point is recreated with simulations in correct azimuth pointings whereas the
2007 paper used simulations pointing to the North.  The updated
correlation plot exhibits the positive correlation between $\gamma$-ray and
hard X-ray fluxes previously reported . However, the reason for such a
correlation remains unclear.  

\section{Summary}
In 2007-2008, VERITAS performed observations on many extragalactic non-blazar
objects such as galaxy cluster, dwarf galaxies, and non-blazar AGNs.  Separate
papers in these proceedings are presented to detail the observations and
preliminary results of Coma cluster \cite{coma} and of Auger hotspots nearby
AGNs \cite{auger}.  

VERITAS spent a total of 55 hours on dwarf galaxies Draco, Ursa Minor, and
Willman I, and established flux upper limits on the order of $1\%$ Crab.  From
Willman I's upper limits, we have derived constraints on the thermally
averaged product of velocity and cross section of WIMP.
 
M87 is detected at $7.08\sigma$ after 41 quality hours, with a marginal TeV
flare that coincided with an X-ray flare in the core detected by Chandra,
which again favors the core as the location of TeV emission. The results from
the multiwavelength observation and the observed rapid variability disflavor
$\gamma$-ray production models that involve the entire galaxy such as the dark
matter annihilation model \cite{baltz99} or the extended kiloparsec jet
\cite{stawarz03}, and strongly suggests the core as the most likely TeV
emission site.


\begin{theacknowledgments}
The authors thank D. E. Harris for providing the Chandra X-ray lightcurves of
M87 nucleus and knot HST-1.

This research is supported by grants from the U.S. Department of Energy, the
U.S. National Science Foundation, and the Smithsonian Institution, by NSERC in
Canada, by PPARC in the UK and by Science Foundation Ireland. 
\end{theacknowledgments}

\end{document}